\documentclass[11pt,a4paper]{article}

\usepackage[utf8]{inputenc}
\usepackage[T1]{fontenc}
\usepackage[expansion=false]{microtype}
\usepackage{amsmath,amssymb,amsthm}
\usepackage{booktabs}
\usepackage{array}
\usepackage{tabularx}
\usepackage{enumitem}
\usepackage{xcolor}
\usepackage{authblk}
\usepackage[margin=2.6cm]{geometry}
\usepackage[colorlinks=true,linkcolor=blue!50!black,citecolor=blue!50!black,urlcolor=blue!50!black]{hyperref}

\theoremstyle{plain}
\newtheorem{theorem}{Theorem}[section]
\newtheorem{lemma}[theorem]{Lemma}
\newtheorem{corollary}[theorem]{Corollary}
\newtheorem{proposition}[theorem]{Proposition}
\theoremstyle{definition}

\newtheorem{assumption}[theorem]{Assumption}
\theoremstyle{remark}
\newtheorem{remark}[theorem]{Remark}

\newcommand{\Adv}{\mathrm{Adv}}

\newcommand{\lsb}{\mathrm{lsb}}
\newcommand{\Ext}{\mathrm{Ext}}
\newcommand{\Sign}{\mathrm{Sign}}
\newcommand{\CDH}{\mathrm{CDH}}
\newcommand{\DLP}{\mathrm{DLP}}
\newcommand{\PRF}{\mathrm{PRF}}
\newcommand{\EUF}{\mathrm{EUF\text{-}CMA}}
\newcommand{\AIPRF}{\mathrm{AI\text{-}PRF}}
\newcommand{\Astar}{A^{*}}
\newcommand{\Z}{\mathbb{Z}}
\newcommand{\bits}{\{0,1\}}
\newcommand{\code}[1]{\texttt{#1}}
\newcommand{\enc}{\mathrm{enc}}
\newcommand{\Pad}{\mathrm{Pad}}
\newcommand{\DLPS}{\mathrm{DLP\text{-}Sign}}

\title{Deterministic ECDSA on secp256k1:\\ the RFC 6979 assumption, a counterexample,\\ and the state of deployed implementations}

\author{Ilia Prosianoi}
\affil{Department of Computer Science, Higher School of Economics, Moscow, Russia\\}
\date{November 2026}

\begin{document}
\maketitle

\begin{abstract}
RFC~6979 derives the ECDSA nonce as $k = F_x(h)$, with $F_x$ HMAC\_DRBG keyed by the signing key $x$. We isolate the assumption its security rests on, $\Astar$: at known public key $Q = xG$, the signing transcript under nonces $F_x(\cdot)$ is indistinguishable from the transcript under uniform nonces. $\Astar$ suffices, and it is neither PRF security of HMAC nor auxiliary-input PRF security ($\AIPRF$) of $F_x$. A function $F'(x,h) = P_{\Ext(xG)}(h)$ is a PRF under standard assumptions yet leaks $x$ from one signature; the same holds for unmodified RFC~6979 over a compression function that keeps the PRF and dual-PRF properties used in proofs of HMAC, given a collision-resistance premise. $\Astar$ and $\AIPRF$ are separated in both directions. $\Astar$ does not imply $\AIPRF$ if the one-bit hidden number problem is hard. Conversely, in the standard model: any $\AIPRF$ can be modified so that it stays one while two chosen-digest signing queries reveal $x$; under DDH with one public random point, or decisional square Diffie--Hellman without it, there is an $\AIPRF$ whose key is recovered from signatures on two ordinary hashed messages; and a third function, under PRF security and DLP, violates $\Astar$ while the scheme stays unforgeable. With an ideal compression function, $\Astar$ follows tightly from DLP in the presence of a randomized-ECDSA signing oracle, and a nonce key independent of $x$ removes the assumption. The secret HMAC message block is $2^{-127.65}$ from uniform on secp256k1, $2^{-32}$ on P-256 and $0.336$ on brainpoolP256r1. An audit of 31 secp256k1 implementations finds that 28 sign deterministically without added entropy and that the raw-message deviation from \code{bits2octets} of libsecp256k1 before 0.2.0 survives in NBitcoin, dcrd and btcec~v2. Low-$R$ grinding needs no assumption beyond $\Astar$, and every nonzero Fourier coefficient of the low-$R$ nonce set is below $2^{-116.9}$ unconditionally. We give test vectors for the untested $h \ge n$ branch.
\end{abstract}

\section{Introduction}

ECDSA leaks the signing key when the per-signature nonce $k$ is repeated, biased, or partially known. Deterministic nonce derivation (RFC~6979~\cite{rfc6979}) removes the dependence on a run-time random number generator by computing $k$ from the signing key $x$ and the message hash $h$ with HMAC\_DRBG. The construction is the de facto standard in the Bitcoin and Ethereum ecosystems: of the 31 secp256k1 implementations we audited, 28 sign ECDSA deterministically by default, and the dominant library, libsecp256k1, offers no other default.

The security argument that accompanies RFC~6979 is informal. Section~4 of the RFC states that HMAC\_DRBG ``is secure in that role as long as HMAC behaves as a PRF'' and that the construction ``requires HMAC\_DRBG to keep on being a random oracle even when the public key is also known''. These two sentences describe two different assumptions, and we show that neither is the one actually needed.

\paragraph{Contributions.}
\begin{enumerate}[leftmargin=*,itemsep=2pt]
\item We formulate the assumption $\Astar$ under which deterministic ECDSA is exactly as secure as randomized ECDSA (Theorem~\ref{thm:astar-suffices}), and show that the auxiliary-input PRF assumption $\AIPRF$ on $F_x$ does not support such a reduction (Section~\ref{sec:reduction}).
\item We give a counterexample: $F'(x,h) = P_{\Ext(xG)}(h)$, where $P$ is any PRF and $\Ext$ extracts $\ell$ bits from the public key, is a PRF under the PRF security of $P$ and a standard randomness-extraction lemma, yet deterministic ECDSA with $F'$ reveals $x$ from one signature (Theorem~\ref{thm:counterexample}). Hence ``PRF + DLP + randomized ECDSA is secure'' does not imply that RFC~6979 is secure, and $\Astar$ is a genuinely additional assumption. We lift the counterexample to the compression function: there is $f^*$ with the PRF and dual-PRF properties used in proofs of HMAC (given a collision-resistance premise) for which the unmodified RFC~6979 state machine is a PRF in $x$ and Det-ECDSA is broken with one signature (Theorem~\ref{thm:whitebox}).
\item We prove that $\Astar$ and $\AIPRF$ are not equivalent. $\Astar \not\Rightarrow \AIPRF$ holds conditionally on the hardness of the one-bit hidden number problem (Section~\ref{sec:nonequiv}). For $\AIPRF \not\Rightarrow \Astar$, the natural candidate---rejecting nonces unless $\lsb(xk \bmod n)=0$---does \emph{not} separate the two notions for any number of signing queries $q$ and any filter width $\ell$ (Theorems~\ref{thm:one-sig} and \ref{thm:hnp}), but two other constructions do, in the standard model. A planted nonce collision turns any $\AIPRF$ into an $\AIPRF$ under which two chosen-digest signing queries reveal $x$ (Theorem~\ref{thm:planted}). Under the decisional square Diffie--Hellman assumption there is a single-key $\AIPRF$ under which signatures on two ordinary SHA-256 messages reveal $x$, so that a third message is forged (Theorem~\ref{thm:dsdh}). Ordinary DDH suffices given one public random point, or, without any setup, given an $\AIPRF$ and about $4\lambda$ signatures; with a second secret scalar the existence of an $\AIPRF$ alone suffices (Theorem~\ref{thm:variants}, Table~\ref{tab:sep}). An ordering bit hidden in a signature scalar violates $\Astar$ with advantage $1/2$ on three fixed hashed messages, under PRF security and DLP, and survives low-$s$ normalization (Theorem~\ref{thm:order}); that scheme remains unforgeable, so $\Astar$ is sufficient but not necessary.
\item We record what is provable: with an ideal compression function, $\Astar$ follows from DLP in the presence of a randomized-ECDSA signing oracle, with no factor $q_H$ (Theorem~\ref{thm:rom}); a nonce key independent of $x$ makes it unnecessary (Section~\ref{sec:provable}). We quantify the curve dependence of the standard part of the assumption (Section~\ref{sec:curves}).
\item We audit 31 implementations (Section~\ref{sec:impl}): nonce defaults, \code{bits2octets} compliance, the retry and grinding logic, and test coverage. We prove that low-$R$ grinding, as deployed in Bitcoin Core, reduces to $\Astar$ on an extended input domain plus randomized ECDSA (Theorem~\ref{thm:grind}), we bound every nonzero Fourier coefficient of the low-$R$ nonce set unconditionally (Proposition~\ref{prop:lowr}), and we supply test vectors for the $h \ge n$ branch, which no implementation we examined tests.
\item We place the one-bit nonce-bias attack in context (Section~\ref{sec:bias}): a modelled data threshold of $2^{47}$ signatures for one error-free leaked bit on a 256-bit curve, with an honest uncertainty envelope of $2^{40}$--$2^{47}$, and the observation that GLV \emph{decomposition} of a uniform nonce creates no such bias.
\end{enumerate}

All claims are tagged in the text as proved here~[T], taken from the literature~[L], established by inspection of source code~[S], or estimates and interpretations~[O].

\section{Preliminaries}\label{sec:prelim}

\paragraph{Curve and notation.}
secp256k1 is the curve $y^2 = x^3 + 7$ over $\mathbb{F}_p$, $p = 2^{256} - 2^{32} - 977$, with prime group order
\[ n = \mathtt{0xFFFFFFFF\,FFFFFFFF\,FFFFFFFF\,FFFFFFFE\,BAAEDCE6\,AF48A03B\,BFD25E8C\,D0364141} \]
and generator $G$. We write $t = 2^{256} - n = 1.2708 \cdot 2^{128}$. The signing key is $x \in [1, n-1]$, the public key is $Q = xG$. For a message $m$ with hash $H(m)$ we write $h = \mathrm{bits2int}(H(m)) \bmod n$. An ECDSA signature on $h$ with nonce $k \in [1,n-1]$ is $(r,s)$ with $R = kG$, $r = x(R) \bmod n$, $s = k^{-1}(h + r x) \bmod n$.

\paragraph{RFC 6979.}
The nonce is $k = F_x(h)$ where
\[ F_x(h) = \mathrm{HMAC\_DRBG}\big(\mathrm{int2octets}(x) \,\|\, \mathrm{bits2octets}(h)\big) \]
and candidates outside $[1,n-1]$ are rejected by advancing the DRBG (step H.3). Section~3.6 of the RFC optionally appends additional data $k'$ to the seed.

\paragraph{Three schemes.}
\emph{Rand-ECDSA}: $k \leftarrow [1,n-1]$ uniformly. \emph{Det-ECDSA}: $k = F_x(h)$. \emph{Det$'$-ECDSA}: $k = P_z(h)$ for a PRF $P$ with key $z$ independent of $x$.

\paragraph{Two assumptions.}
Let $D$ be a probabilistic polynomial-time distinguisher and $\mathcal{R}$ a random function.

\begin{assumption}[$\AIPRF$]\label{ass:aiprf}
$F$ is an auxiliary-input PRF at the public key if for every $D$
\[
\Big|\Pr\big[D^{F_x(\cdot)}(Q) = 1\big] - \Pr\big[D^{\mathcal{R}(\cdot)}(Q) = 1\big]\Big| \le \varepsilon .
\]
\end{assumption}

\begin{assumption}[$\Astar$]\label{ass:astar}
$F$ satisfies $\Astar$ if for every $D$
\[
\Big|\Pr\big[D^{\Sign_x(\cdot;\,F_x)}(Q) = 1\big] - \Pr\big[D^{\Sign_x(\cdot;\,\mathcal{R})}(Q) = 1\big]\Big| \le \varepsilon ,
\]
where $\Sign_x(\cdot; F_x)$ is the signing oracle using nonces $F_x(h)$ and $\Sign_x(\cdot;\mathcal{R})$ the oracle using independent uniform nonces (with the same rejection rule).
\end{assumption}

Both assumptions are falsifiable in the sense of Naor~\cite{naor2003}: an efficient distinguisher is a verifiable witness of failure.

\paragraph{Chosen digests and hashed messages.}
In both assumptions the distinguisher submits digests $h \in \Z_n$ directly; we call this the \emph{chosen-digest} game. It is the interface of libsecp256k1 and of most libraries of Section~\ref{sec:impl}, which sign a caller-supplied 32-byte digest. In the \emph{hashed-message} game queries are restricted to $h = \mathrm{bits2int}(H(m)) \bmod n$ for messages $m$. An assumption stated for chosen digests implies its hashed-message version, whereas an attack that needs particular digests may not transfer. Forgery games are always played with hashed messages, because ECDSA is forgeable on chosen digests (see~\eqref{eq:sim}).

\section{What the reduction needs}\label{sec:reduction}

\begin{theorem}[$\Astar$ suffices]\label{thm:astar-suffices}
For every forger $A$ against Det-ECDSA there is a forger $B$ against Rand-ECDSA and a distinguisher $D$ with
\[
\Adv^{\EUF}_{\mathrm{Det}}(A) \le \Adv^{\EUF}_{\mathrm{Rand}}(B) + \Adv_{\Astar}(D).
\]
\end{theorem}
\begin{proof}
Game~0 is the Det-ECDSA forgery game; Game~1 replaces the nonces by independent uniform values. Any $A$ whose success differs between the games is a distinguisher $D$ for $\Astar$: $D$ receives $Q$, forwards signing queries to its oracle, and outputs~1 iff $A$ forges. Game~1 is the Rand-ECDSA game. [T]
\end{proof}

\begin{proposition}[$\AIPRF$ does not support the reduction]\label{prop:aiprf-insufficient}
A distinguisher for $\AIPRF$ receives $(Q, k_1, \dots, k_q)$. To simulate the signing oracle for a Det-ECDSA forger it must produce $s_i = k_i^{-1}(h_i + r_i x)$, which requires $x$. No generic reduction from $\AIPRF$ to the security of Det-ECDSA can simulate signatures in the standard model. [T]
\end{proposition}

The proposition says only that the \emph{obvious} reduction fails. The next section shows that \emph{no} reduction that uses $F_x$ purely as a PRF can succeed, and Theorems~\ref{thm:planted} and~\ref{thm:dsdh} that the same is true of $\AIPRF$: there are nonce functions satisfying it under which two signatures reveal the key, on chosen digests if any $\AIPRF$ exists at all, and on ordinary hashed messages under decisional square Diffie--Hellman, or under DDH with a public random point.

\section{A counterexample}\label{sec:counterexample}

Let $\enc$ be an injective encoding of the points of $\langle G\rangle$, including $\mathcal{O}$, as strings of at most 258 bits, and let $\{h_s\}_s$ be a universal family of functions $\bits^{258} \to \bits^{\ell}$. For a seed $s$ fixed below put $\Ext(Q) := h_s(\enc(Q))$. Let $P$ be any PRF with a uniform $\ell$-bit key, $\ell = 128$. Define
\begin{equation}\label{eq:fprime}
F'(x, h) := P_{\Ext(xG)}(h).
\end{equation}
The key $x$ enters $F'$ only through $\Ext(xG)$, which is a function of the public key.

\begin{lemma}[extraction]\label{lem:ext}
For $x$ uniform on $[1,n-1]$,
\[ \mathbb{E}_s\,\mathrm{SD}\big(h_s(\enc(xG)),\,U_\ell\big) \le \tfrac12\, 2^{(\ell - \log_2(n-1))/2}, \]
which is $2^{-65}$ for $\ell = 128$. Hence there is a seed $s$ with $\delta_{\Ext} := \mathrm{SD}(\Ext(xG), U_\ell) \le 2^{-65}$; we fix such an $s$. [T]
\end{lemma}
\begin{proof}
$xG$ is uniform on $n-1$ points and $\enc$ is injective, so $H_\infty(\enc(xG)) = \log_2(n-1)$. Apply the leftover hash lemma~\cite{ill1989} and average over $s$.
\end{proof}

\begin{remark}[deterministic truncation]
A seedless alternative is to truncate the abscissa, but the available bounds are for the \emph{least} significant bits: Chevalier, Fouque, Pointcheval and Zimmer~\cite[Thm.~14]{cfpz2009} prove $\mathrm{SD}(\lsb_k(x(U)), U_k) \le 2^{(k + m + \log_2 m)/2 + 3 - \log_2 n}$ for $U$ uniform on $\langle G\rangle$ and $m = \lceil \log_2 p\rceil$, i.e.\ $2^{k/2-121}$ on secp256k1 and $2^{-57}$ for $k = 128$. The work of Fouque, Pointcheval, Stern and Zimmer~\cite{fpsz2006} concerns subgroups of $\Z_p^*$ only. We use the seeded extractor throughout because Section~\ref{sec:whitebox} needs extraction from a non-uniform subset of the group, where the leftover hash lemma applies verbatim.
\end{remark}

\begin{lemma}[PRF security]\label{lem:fprime-prf}
$\Adv^{\PRF}_{F'}(A) \le \Adv^{\PRF}_{P}(B) + \delta_{\Ext}$. [T]
\end{lemma}
\begin{proof}
Replace $\Ext(xG)$ by a uniform key (loss $\delta_{\Ext}$ by Lemma~\ref{lem:ext}); $B$ forwards $A$'s queries to its $P$ oracle. No circular-security assumption is needed because $x$ is used only as PRF key material.
\end{proof}

\begin{lemma}[key recovery]\label{lem:fprime-break}
Given $Q$ and one Det-ECDSA signature $(r,s)$ on $h$ with $k = F'(x,h)$, anyone computes $K = \Ext(Q)$, $k = P_K(h)$ and $x = (sk - h)\, r^{-1} \bmod n$. [T]
\end{lemma}

\begin{theorem}[counterexample]\label{thm:counterexample}
Assume $P$ is a PRF, DLP is hard on secp256k1 and Rand-ECDSA is $\EUF$-secure. Then $F'$ is a PRF (Lemma~\ref{lem:fprime-prf}), all three premises hold, and Det-ECDSA with $F'$ is totally broken (Lemma~\ref{lem:fprime-break}). Consequently no reduction that treats the nonce function only as a PRF---in particular none that uses only the PRF security of HMAC---can prove RFC~6979 secure. $F'$ also violates $\AIPRF$: anyone holding $Q$ recomputes all its outputs. [T]
\end{theorem}

\begin{remark}
A variant $F^{\circ}(x,h) = x + P_{\Ext(xG)}(h)$ would require adding $x$ to oracle answers in the PRF reduction and therefore a circular-key assumption; the variant~\eqref{eq:fprime} avoids this entirely.
\end{remark}

\subsection{Opening the box: the SHA-256 compression function}\label{sec:whitebox}

Theorem~\ref{thm:counterexample} treats $F_x$ as a black box. HMAC\_DRBG is not an arbitrary PRF: proofs of HMAC rest on properties of the SHA-256 compression function $f : \bits^{256} \times \bits^{512} \to \bits^{256}$, and one might hope that these properties, together with the specific state machine of RFC~6979~\S3.2, imply $\Astar$. They do not, given a collision-resistance premise.

\paragraph{Block layout.} With $K_0 = \mathtt{0x00}^{32}$ and $V_0 = \mathtt{0x01}^{32}$, step~d computes
\[ K_1 = \mathrm{HMAC}_{K_0}\big(V_0 \,\|\, \mathtt{0x00} \,\|\, \mathrm{int2octets}(x) \,\|\, \mathrm{bits2octets}(h)\big). \]
The inner hash first compresses the public block $(K_0\|0^{256}) \oplus \mathrm{ipad} = \mathtt{0x36}^{64}$, giving a public chaining value $IV_1$, and then the 97-byte message; the total length is 161 bytes, $L = 1288$ bits. The two message blocks are
\[
m_1 = V_0 \,\|\, \mathtt{0x00} \,\|\, x_{[0..30]}, \qquad
m_2 = x_{[31]} \,\|\, \mathrm{bits2octets}(h) \,\|\, \mathtt{0x80} \,\|\, 0^{176} \,\|\, \langle L\rangle_{64},
\]
where $x_{[i]}$ is the $i$-th byte of $\mathrm{int2octets}(x)$. Step~f has the same layout with $V_1$, $\mathtt{0x01}$ and the secret chaining value derived from $K_1$. Write $\Pad = \{\mathtt{0x80}\|0^{176}\|\langle L\rangle_{64} : L \ne 0\}$ for the 31-byte SHA-256 padding tails. In every other block of the computation---HMAC key setup, outer hashes, steps e, g, h, and the retry step H.3---either byte~32 is not in $\{\mathtt{0x00},\mathtt{0x01}\}$ or bytes 33--63 lie in $\Pad$. Writing $x = 2^8\tilde x + b$ with $\tilde x = \mathrm{int}(x_{[0..30]})$ and $b = x_{[31]}$, we have $2^8 \tilde x G = Q - bG$. [T]

\paragraph{Properties of $f$.} We use three properties.
\begin{enumerate}[leftmargin=*,itemsep=1pt]
\item[(A1)] \emph{PRF in the chaining value}: $f(cv,\cdot)$ is a PRF for uniform secret $cv$.
\item[(A2)] \emph{Dual PRF with a key in the first half of the block}: for uniform secret $u \in \bits^{256}$, the map $(cv, w) \mapsto f(cv, u\|w)$ is a PRF. This is the property used for HMAC with 256-bit keys and for the outer hash~\cite{bellare2006,bbgs2023}.
\item[(A2$'_\lambda$)] \emph{Tail-keyed pseudorandomness}: for public $cv_0$, a public 264-bit prefix $\pi$ and uniform $K \in \bits^\lambda$, the value $y = f(cv_0,\, \pi\|K\|0^{248-\lambda})$ is pseudorandom. A standard-model proof that RFC~6979 is a PRF in $x$ needs (A2$'_{248}$) with $\pi = V_0\|\mathtt{0x00}$ and $K = x_{[0..30]}$ (Section~\ref{sec:curves}).
\end{enumerate}

\paragraph{Construction.} Let $\ell \le 184$, let $\{h_s\}$ be as in Section~\ref{sec:counterexample} and read $w \in \bits^{248}$ as an integer. Define
\[
f^*(cv,\ u\|c\|w) =
\begin{cases}
f\big(cv,\ u \,\|\, c \,\|\, h_s(\enc(2^8 w G)) \,\|\, 0^{248-\ell}\big), & c \in \{\mathtt{0x00},\mathtt{0x01}\},\ w \notin \Pad,\\[2pt]
f(cv,\ u\|c\|w), & \text{otherwise,}
\end{cases}
\]
with $u \in \bits^{256}$ and $c \in \bits^8$. Let $\Adv^{\mathrm{CR}}$ denote the advantage of an efficient algorithm that, given $s$, outputs $w \ne w'$ in $\bits^{248}\setminus\Pad$ with $h_s(\enc(2^8wG)) = h_s(\enc(2^8w'G))$, and put $\delta_\ell = \frac12\,2^{(\ell-248)/2}$.

\begin{lemma}[$f^*$ inherits the properties]\label{lem:fstar}
$\Adv^{\mathrm{A1}}_{f^*} \le \Adv^{\mathrm{A1}}_f + \Adv^{\mathrm{CR}}$, $\Adv^{\mathrm{A2}}_{f^*} \le \Adv^{\mathrm{A2}}_f + \Adv^{\mathrm{CR}}$, and $f^*$ satisfies (A2$'_{248}$) whenever $f$ satisfies (A2$'_{\ell}$) and (A2$'_{248}$), with additional loss $\delta_\ell + 2^{-184}$. [T]
\end{lemma}
\begin{proof}
$f^*(cv, m) = f(cv, \varphi(m))$ for a public map $\varphi$. The two branches never collide under $\varphi$: on the first branch the last $248-\ell \ge 64$ bits are zero, whereas every element of $\Pad$ ends in a non-zero length field, and for $c \notin \{\mathtt{0x00},\mathtt{0x01}\}$ the byte $c$ already differs. Hence $\varphi$ is injective on any set of queries that contains no collision of the composite map, and a distinguisher for (A1) or (A2) of $f^*$ is one for $f$ unless it finds such a collision. For (A2$'_{248}$) with prefix byte $c \in \{\mathtt{0x00},\mathtt{0x01}\}$ and uniform $w$: $w \in \Pad$ with probability $\le 2^{64-248}$; the map $w \mapsto 2^8wG$ is injective on $[0,2^{248})$ because $2^8$ is invertible modulo $n$ and $|w-w'| < n$, so $H_\infty(\enc(2^8wG)) = 248$ and the leftover hash lemma gives distance $\delta_\ell$ for all but a small fraction of seeds; choose $s$ good for this and for Lemma~\ref{lem:ext} simultaneously. Then apply (A2$'_\ell$). For other prefix bytes $f^* = f$.
\end{proof}

\begin{theorem}[white-box counterexample]\label{thm:whitebox}
Assume $f$ satisfies (A1), (A2) and (A2$'_\ell$) and that the composite map above is collision resistant. Let $F^*_x(h)$ be the unmodified RFC~6979~\S3.2 nonce derivation with the SHA-256 compression function replaced by $f^*$. Then:
\begin{enumerate}[leftmargin=*,itemsep=1pt]
\item $f^*$ satisfies (A1), (A2) and (A2$'_{248}$) (Lemma~\ref{lem:fstar});
\item $F^*$ is a PRF in $x$: for $q$ queries, $\Adv^{\PRF}_{F^*} \le \delta_\ell + 256/n + 2^{-184} + \Adv^{\mathrm{A2}'_\ell}_f + O(q)\big(\Adv^{\mathrm{A1}}_f + \Adv^{\mathrm{A2}}_f\big)$;
\item given $Q$ and one Det-ECDSA signature under $F^*$, $x$ is recovered with at most 256 runs of the nonce derivation, so Det-ECDSA with $F^*$ is totally broken and $\Astar$ fails.
\end{enumerate}
Consequently $\Astar$ for HMAC\_DRBG does not follow from (A1), (A2) and (A2$'_{248}$) of the compression function together with DLP and the EUF-CMA security of Rand-ECDSA, provided a collision-resistant map of this form exists. [T]
\end{theorem}
\begin{proof}
(2) The pair $(\tilde x, b)$ is within $256/n$ of independent uniform values on its range. $\tilde x$ enters the computation only in block~1 of steps d and f, and there only through $K = h_s(\enc(2^8\tilde xG))$, except when $\tilde x \in \Pad$. Replace $K$ by a uniform $U$ (loss $\delta_\ell$); $\tilde x$ then disappears from the game. Now $cv_1 = f(IV_1, V_0\|\mathtt{0x00}\|U\|0^{248-\ell})$ is pseudorandom by (A2$'_\ell$). Block~2 of step~d is $f(cv_1, b\|h\|\cdots)$, a random function of $h$ by (A1); distinct queries $h$ give distinct inputs. The outer hash and the key setup of $K_1$ are keyed in the first half of the block by (A2); step~e is (A1); block~1 of step~f is $f(K_1^{\mathrm{in}}, V_1\|\mathtt{0x01}\|U\|0^{248-\ell})$ under (A1); the remaining evaluations of steps f--h and the retries are evaluations of $f$ handled by (A1) and (A2). The final nonces are independent and uniform for distinct $h$.

(3) For each $\hat b \in [0,255]$ compute $\hat K = h_s(\enc(Q - \hat bG))$. Every evaluation of $f^*$ in the derivation depends on $x_{[0..30]}$ only through $2^8\tilde xG$, hence through $\hat K$ when $\hat b = b$, and on $x_{[31]}$ directly. Run the derivation with $\hat K$ and $\hat b$, obtain $\hat k$, and test $\hat x = (s\hat k - h)r^{-1}$ against $\hat xG = Q$. For $\hat b = b$ this returns $x$, unless $\tilde x \in \Pad$ (probability $\le 2^{-184}$).
\end{proof}

\begin{table}[h]
\centering
\caption{Parameters of $f^*$: extraction loss and generic cost of a collision. [T][O]}\label{tab:fstar}
\begin{tabular}{@{}lll@{}}
\toprule
$\ell$ & $\delta_\ell$ & collision, generic group \\
\midrule
128 & $2^{-61}$ & $2^{64}$ \\
160 & $2^{-45}$ & $2^{80}$ \\
184 & $2^{-33}$ & $2^{92}$ \\
\bottomrule
\end{tabular}
\end{table}

\begin{remark}[on the collision premise]
$h_s$ alone is not collision resistant once $s$ is known: a linear universal family has collisions in a known kernel. Exploiting a kernel element, however, requires a $w$ with prescribed point $2^8wG$, i.e.\ a discrete logarithm. In the generic group model~\cite{shoup1997} encodings are random and collisions of the composite cost $2^{\ell/2}$ (Table~\ref{tab:fstar}). We know no reduction to DLP, so the premise is non-standard and is stated explicitly. [O]
\end{remark}

\begin{remark}[why the collision term is there]
A proof of item~(2) by statistical extraction must replace all information about $\tilde x$ that enters $f^*$ by a uniform string, which the leftover hash lemma allows only for strings shorter than $H_\infty(\tilde x) = 248$ bits. The first branch of $f^*$ is then not injective in $w$, and (A1) for $f^*$ needs collision resistance. A two-call variant that feeds $\enc(2^8wG)$ in clear into a second compression call restores injectivity but breaks item~(2): after $K$ is replaced by $U$, the point still appears in the second call of steps d and f, and the hybrid argument cannot proceed. Computational extraction could avoid the term at the price of another non-standard assumption. Theorem~\ref{thm:whitebox} does not exclude proofs of $\Astar$ from other properties of $f$, such as correlation intractability, or in the random-oracle model (Theorem~\ref{thm:rom}). [O]
\end{remark}

\section{\texorpdfstring{$\Astar$}{A*} and \texorpdfstring{$\AIPRF$}{AI-PRF} are not equivalent}\label{sec:nonequiv}

\subsection{\texorpdfstring{$\Astar \not\Rightarrow \AIPRF$}{A* does not imply AI-PRF}}\label{sec:astar-not-aiprf}

Take $F$ with a one-bit leak: $F_x(h) = P_x(h)$ with its least significant bit replaced by a bit computable from $(x,h)$ and checkable from $Q$. $\AIPRF$ is broken with $O(1)$ queries. For $\Astar$ an observer of signatures knows one bit of every $k_i$; this is the hidden number problem with one known bit per sample, equivalently a half-range nonce after the substitution $k = 2k' + b$. The best known attack is Bleichenbacher's at $2^{N/5}$ (Aranha et al.~\cite{aranha2014}), about $2^{52}$ for 256 bits by scaling, $2^{47}$ by the model of Section~\ref{sec:bias}. A generic reduction $\AIPRF$-distinguisher $\Rightarrow$ $\Astar$-distinguisher would therefore yield a polynomial-time distinguisher for signatures with one-bit nonce bias. The separation holds conditionally on the one-bit HNP being hard, which is the same assumption the practical relevance of RFC~6979 rests on. [O]

\subsection{\texorpdfstring{$\AIPRF \not\Rightarrow \Astar$}{AI-PRF does not imply A*}: a natural candidate that fails}\label{sec:fdprime}

A separating predicate $\pi(x,k)$ must (i)~be computable by the signer, who knows $x$; (ii)~not be computable from $(Q,k)$, else the $\AIPRF$ distinguisher sees it; (iii)~be efficiently detectable from the transcript $(Q, R, s, h)$, else $\Astar$ is not violated. The natural candidate is $\pi(x,k) = \lsb(xk \bmod n)$. (Conditions on the point $kQ$ fail~(ii): the $\AIPRF$ distinguisher computes $kQ$ itself.)

\paragraph{Construction.} Let $P$ be a PRF with an independent key $z$ and define
\begin{equation}\label{eq:fdprime}
F''_{(x,z)}(h) := P_z(h, i^*), \qquad i^* = \min\{\, i : \lsb_\ell(x \cdot P_z(h,i) \bmod n) = 0^\ell \,\}.
\end{equation}
The expected number of trials is $2^\ell$. The PRF key must not be $x$ itself, since the $\AIPRF$ security of $P_x$ at known $Q$ is the assumption under study.

\paragraph{Simulatability of a signature.}
For uniform $h$ and $k$, the tuple $(Q,R,s,h)$ has the same distribution as the tuple built without $x$:
\begin{equation}\label{eq:sim}
a \leftarrow \Z_n,\ b \leftarrow \Z_n^*,\ R = aG + bQ,\ r = x(R) \bmod n,\ s = r b^{-1},\ h = a r b^{-1},
\end{equation}
under which $k = a + bx$ and $xk \equiv ax + bx^2 \pmod n$; the inverse map is $a = s^{-1}h$, $b = s^{-1}r$. This is the observation underlying Fersch, Kiltz and Poettering~\cite{fkp2016}. Consequently an observer of one signature solves ``$\lsb(ax + bx^2 \bmod n)$ from $(Q,a,b)$'' with uniform $(a,b)$. [T]

\begin{theorem}[one signature]\label{thm:one-sig}
A predictor of $\lsb(xk \bmod n)$ from $(Q,R,s,h)$ with advantage $\delta$ over $1/2$, for uniform $h$ and $k$, yields a DLP solver running in time $\mathrm{poly}(\log n, 1/\delta)$. [T]
\end{theorem}
\begin{proof}
By~\eqref{eq:sim} the reduction constructs the tuple for any chosen $(a,b)$ and so obtains an oracle $\tilde B(a,b)$ predicting $\lsb(ax + bx^2 \bmod n)$. By Markov's inequality there is $b_0$ for which $g(a) := \tilde B(a,b_0)$ predicts $\lsb([ax + c_0]_n)$, $c_0 = b_0 x^2$, with advantage $\ge \delta/2$. Lemma~\ref{lem:acgs} recovers $x$.
\end{proof}

\begin{lemma}[ACGS with an unknown shift]\label{lem:acgs}
Let $x \in \Z_n^*$, let $c_0 \in \Z_n$ be unknown, and let $g : \Z_n \to \bits$ satisfy $\Pr_a[g(a) = \lsb([ax + c_0]_n)] \ge \tfrac12 + \varepsilon$. Then $O(\varepsilon^{-2}\log n)$ non-adaptive queries to $g$ and $O(\varepsilon^{-5} \log n)$ time recover $x$ with probability $\ge 3/4$. [T]
\end{lemma}
\begin{proof}[Proof sketch]
Follow Alexi, Chor, Goldreich and Schnorr~\cite{acgs1988} (in the form of Fischlin and Schnorr~\cite{fs2000}). Query points $a_{k,i} = 2^{-k} + iu + v$ for random $(u,v)$, $k = 0,\dots,\lceil \log_2 n \rceil$, $i = 1,\dots,m$. Then $a_{k,i}x + c_0 \equiv 2^{-k}x + iY + Z$ with $Y = [ux]_n$ and $Z = [vx + c_0]_n$. The unknown shift enters only through $Z$, whose least significant bit and coarse approximation are enumerated exactly as ACGS enumerates them for $[vx]_n$; the search space does not grow. The $a_{k,i}$ are pairwise independent in $i$ because $n$ is prime, so Chebyshev's inequality bounds the error of the majority vote at each halving step; a union bound over the $\lceil \log_2 n\rceil + 1$ steps gives success probability $3/4$ with $m = \lceil 8 \log_2 n / \varepsilon^2\rceil$. The final candidate $2^{K}\lceil n\gamma_K \rceil \bmod n$ is checked against $xG = Q$. Full details are in Appendix~\ref{app:acgs}.
\end{proof}

\begin{corollary}[$q=1$]\label{cor:q1}
With one signing query, $F''$ with $\ell = 1$ satisfies both $\AIPRF$ (Yao's lemma plus the hardcore property of the least significant bit of the discrete logarithm~\cite{hn2004}; no hybrid is needed) and $\Astar$ (Theorem~\ref{thm:one-sig}), assuming $P$ is a PRF and DLP is hard. It does not separate the notions. For $\ell > 1$ the simultaneous hardcore property of $\ell$ bits is needed. [T][L]
\end{corollary}

\begin{theorem}[$q > 1$]\label{thm:hnp}
For $q > 1$ signing queries, neither side of $F''$ reduces to DLP:
\begin{enumerate}[leftmargin=*,itemsep=1pt]
\item The hybrid argument fails for $\AIPRF$: Yao's lemma over $q$ samples with a common secret $x$ requires the reduction to sample $k_i \in S_x = \{k : \lsb(xk \bmod n) = 0\}$ without knowing $x$, which is the bit it is trying to predict. The $\AIPRF$ game for $F''$ is the one-dimensional HNP
\[ (2^{-\ell}k_i)\,x \bmod n \in [0, n/2^\ell), \qquad i = 1,\dots,q .\]
\item On the $\Astar$ side the hint $s_i$ eliminates $k_i$: $xk_i \equiv a_i x + b_i y$ with $y = x^2$, and the filter gives the two-dimensional HNP
\[ (2^{-\ell}a_i)\,x + (2^{-\ell}b_i)\,y \bmod n \in [0, n/2^\ell), \qquad i = 1,\dots,q .\]
\end{enumerate}
For $\ell \ge 4$ on a 256-bit curve both are solved by the Boneh--Venkatesan lattice~\cite{bv1996}, the $\AIPRF$ side in lower dimension and with half the samples. For $\ell = 1$ both are conditional on the hardness of one-bit HNP, in one and two dimensions respectively. A separation would require ``1D-HNP hard, 2D-HNP easy'', for which there is no evidence. [T]
\end{theorem}

\begin{table}[t]
\centering
\caption{Status of the candidate $F''$ by number of queries $q$ and filter width $\ell$.}\label{tab:fdprime}
\small
\begin{tabularx}{\textwidth}{@{}lXXl@{}}
\toprule
Regime & $\AIPRF$ & $\Astar$ & Separation \\
\midrule
$q = 1$, $\ell = 1$ & holds under PRF + DLP [T] & holds under PRF + DLP (Thm.~\ref{thm:one-sig}) [T] & no \\
$q > 1$, $\ell \ge 4$ & broken: 1D-HNP, lattice [T] & broken: 2D-HNP, lattice [T] & no \\
$q > 1$, $\ell = 1$ & conditional on 1-bit 1D-HNP & conditional on 1-bit 2D-HNP & no \\
\bottomrule
\end{tabularx}
\end{table}

\begin{remark}[oracle separation]\label{rem:oracle}
$F''$ does separate the notions relative to an oracle. Let $H$ be a random oracle and let $\mathcal{O}_Q(R, s, m)$ return $\lsb(\CDH(Q,R))$ when $r = x(R) \bmod n \ne 0$, $s \in \Z_n^*$ and $sR = H(m)G + rQ$, and $\bot$ otherwise. An observer of one signature then breaks $\Astar$ with advantage $1/2$, while $\AIPRF$ at $q = 1$ holds up to the probability of a key-only forgery under $Q$, which reduces to DLP in the bijective random-oracle model of~\cite{fkp2016}: the reduction answers $\bot$ throughout, and the hash-free forgery $R = aG + bQ$ does not qualify because $h = ar/b \ne H(m)$. The oracle must be bound to $Q$; one that accepts arbitrary keys $Q^*$ admits a rogue-key attack ($R := Q$, $s := 1$, $Q^* := r^{-1}(R - H(m)G)$). Section~\ref{sec:sep} makes the oracle unnecessary. [T]
\end{remark}

\begin{remark}
Candidates of the type of $F''$ fail for a simple reason: a relation between the nonce $k_i$ and $x$ that the signature equation $s_i k_i = h_i + r_i x$ exposes is exposed more directly by the public pair $(Q, k_i)$ in the $\AIPRF$ game. [O]
\end{remark}

\subsection{\texorpdfstring{$\AIPRF \not\Rightarrow \Astar$}{AI-PRF does not imply A*}: standard-model separations}\label{sec:sep}

$F''$ places the relation between the nonce and $x$ in the \emph{output} of the nonce function, where the $\AIPRF$ adversary reads it off. The three constructions below hide it elsewhere: in an input that only a signature reveals, in a PRF key derived from a point that only a signature reveals, and in one bit of a signature scalar, which a nonce observer knows only in the exponent. Write $r(k) = x(kG) \bmod n$. Since $p < 2n$, at most two field elements reduce to a given $r$ and each is the abscissa of at most two points, so at most four scalars $k$ share a value $r(k)$.

\begin{theorem}[planted collision: one key, two chosen digests]\label{thm:planted}
Let $P$ be a nonce function with outputs in $\Z_n^*$ that satisfies $\AIPRF$ at $Q = xG$. Put $a = P_x(0)$ and
\[
F^{\dagger}_x(h) =
\begin{cases}
a, & ah \in \{x, -x\},\\
P_x(h), & \text{otherwise.}
\end{cases}
\]
\begin{enumerate}[leftmargin=*,itemsep=1pt]
\item For every distinguisher $D$ there are $B$ and $C$, running in the time of $D$ plus one scalar multiplication per query, with
\[
\Adv^{\AIPRF}_{F^\dagger}(D) \le \Adv^{\AIPRF}_{P}(B) + \Pr[C(xG) = x].
\]
There is no factor in the number of queries.
\item There is a distinguisher $D^\dagger$ making two signing queries with
\[
\Adv_{\Astar}(D^\dagger) \ge 1 - \Adv^{\AIPRF}_{P}(B') - 8/(n-9),
\]
also against a signer that normalizes to low $s$. The two signatures reveal $x$.
\end{enumerate}
[T]
\end{theorem}
\begin{proof}
(1) $B$ receives $Q$ and an oracle $O$, asks $a = O(0)$, and answers a query $h$ of $D$ with $a$ if $(ah)G \in \{Q, -Q\}$ and with $O(h)$ otherwise. For $O = P_x$ this is exactly $F^\dagger_x$. For a random function $O = \mathcal{R}$ the answers are those of $\mathcal{R}$ unless $D$ submits an exceptional $h$; $C$ runs the same simulation with a lazily sampled $\mathcal{R}$ and, at the first exceptional query, outputs $ah$ or $-ah$ according to which point matched. The simulator recognizes the hit itself, so no query index is guessed.

(2) $D^\dagger$ queries the digest $0$, receives $(r_0, s_0)$ and computes $t = s_0/r_0$. The raw signature scalar is $a^{-1} r_0 x$, so $t = \pm x/a$, the sign depending on normalization; this is why both signs are exceptional. $D^\dagger$ picks $h \in \{t, -t\}$ with $hG + r_0Q \ne \mathcal{O}$, which holds for at least one sign because $t \ne 0$, queries $h$, and outputs~1 iff $r_1 = r_0$. In the real game $ah = \pm x$, so the nonce is $a$ again and $r_1 = r_0$; the choice of sign makes $s_1 \ne 0$, so there is no retry. The real game fails only if $r(a) = 0$, an event that is publicly testable from $a = O(0)$ and therefore has probability at most $4/(n-1) + \Adv^{\AIPRF}_P(B')$. In the ideal game $h \ne 0$ is fresh and the nonce is uniform among the at least $n - 9$ valid candidates, of which at most four give $r_1 = r_0$.

For key recovery, the raw scalars satisfy $a s_0 = r_0 x$ and $a s_1 = h + r_0 x$, hence $a = h/(s_1 - s_0)$ and $x = a s_0 / r_0$; under low-$s$ normalization the four sign combinations are tried and tested against $Q$.
\end{proof}

\begin{remark}[scope of Theorem~\ref{thm:planted}]
A discrete-logarithm solver distinguishes $P_x$ from a random function at known $Q$, so $\Pr[C(xG) = x] \le 2\,\Adv^{\AIPRF}_P(C')$: the theorem assumes only that some $\AIPRF$ exists, and turns it into one under which signing leaks the key. The secret $x$ is the only key; no independent seed is used. The attack needs digests of the adversary's choice, namely $0$ and $\pm x/a$. The anchor $0$ can be replaced by any fixed digest $h_0$ by planting the collision at $ah = \pm(h_0 + r(a)x)$, that is at $h = \pm s_0$, which is tested publicly by $(ah)G = \pm(h_0G + r(a)Q)$. The second digest still depends on the first signature, so the attack does not pass through a hash function without computing preimages; Theorem~\ref{thm:dsdh} removes this restriction at the price of a stronger assumption. No comparable attack on Rand-ECDSA is known: key recovery with a chosen-digest signing oracle is the assumption $\DLPS$ of Theorem~\ref{thm:rom}. [T]
\end{remark}

\begin{theorem}[hidden seed: one key, two hashed messages]\label{thm:dsdh}
Let $\lambda = \lceil \log_2 p \rceil$ and $\kappa = \lfloor \lambda/4 \rfloor$, let $P$ be a PRF with $\kappa$-bit keys and outputs in $\Z_n^*$, and let $m_0$ be a fixed public message with digest $h_0$. Fix a representation of the field $\mathbb{F}_{2^\lambda}$ and write $\odot$ for its multiplication, applied to abscissas read as $\lambda$-bit strings (the point at infinity is encoded by $p$). For a signing key $x$ with $Q = xG$ let
\[
W = x^2G,\qquad J = \Big\lfloor \frac{2^\kappa\, x(W)}{p} \Big\rfloor,\qquad a = P_J(h_0),\qquad c = \frac{h_0}{r(a)},\qquad T = a^{-1}\big(W - c^2G\big),
\]
with $c = 0$ if $r(a) = 0$, let $L = \mathrm{msb}_\kappa\big(x(Q) \odot x(T)\big)$, and define
\[
F^{\sharp}_x(h) =
\begin{cases}
P_J(h), & h = h_0,\\
P_L(h), & h \ne h_0.
\end{cases}
\]
Candidates rejected by the signer are replaced by including a retry counter in the PRF input. On secp256k1, where $\kappa = 64$:
\begin{enumerate}[leftmargin=*,itemsep=1pt]
\item for every distinguisher $D$, making any number of adaptive queries,
\[
\Adv^{\AIPRF}_{F^\sharp}(D) \le \Adv^{\mathrm{DSDH}}(B) + \Adv^{\PRF}_{P}(B_1) + \Adv^{\PRF}_{P}(B_2) + 2^{-55.9},
\]
where $\Adv^{\mathrm{DSDH}}(B)$ is the advantage of $B$ in distinguishing $(Q, x^2G)$ from $(Q, W)$ with $W$ a uniform point independent of $Q$;
\item an adversary that sees $Q$ and the signatures on $m_0$ and on one further message $m_1$ of its choice recovers $x$, except with probability at most $\Adv^{\AIPRF}_{F^\sharp}(D') + 8/(n-1)$, and then signs any message. This holds in the hashed-message game and against a signer that normalizes to low $s$.
\end{enumerate}
[T]
\end{theorem}
\begin{proof}
(1) The whole nonce oracle is a function of the two points $(Q, W)$: $B$ computes $J$, $a$, $c$, $T$ and $L$ from them without knowing $x$ and answers every query. For $W = x^2G$ this is $F^\sharp_x$, so replacing $W$ by an independent uniform point costs $\Adv^{\mathrm{DSDH}}(B)$, for adaptive queries on arbitrary digests. In that hybrid two statistical facts suffice.

\emph{$J$ is close to uniform.} The value $J = j$ corresponds to an interval of abscissas of length $p/2^\kappa$. The character-sum estimate in the proof of Proposition~\ref{prop:lowr} applies to every interval, so the number $N_j$ of points in the $j$-th bucket differs from $n/2^\kappa$ by at most $D + 3$ with $D < 2^{137.06}$, and the statistical distance of $J$ from uniform is at most $2^{\kappa - 1}(D+3)/n < 2^{-55.9}$.

\emph{$L$ is close to uniform given $Q$ and $J$.} Condition on $J = j$. Then $a$ and $c$ are fixed, $W$ is uniform on $N_j$ points and $W \mapsto a^{-1}(W - c^2G)$ is a bijection, so $T$ is uniform on $N_j$ points. An abscissa has at most two preimages, so the collision probability of $x(T)$ is at most $2/N_j$, and on average over $j$ at most $\sum_j (N_j/n)(2/N_j) \le 2^{\kappa+1}/n$. For a uniform multiplier $u$ the family $v \mapsto \mathrm{msb}_\kappa(u \odot v)$ is universal, and the average-case leftover hash lemma~\cite{dors2008} bounds the distance of $L$ from uniform by $\tfrac12\sqrt{2^{2\kappa+1}/n}$. The actual multiplier $x(Q)$ is independent of $(J, T)$ and takes each value with probability at most $2/(n-1)$, which is at most $2^{\lambda+1}/(n-1) < 2.01$ times the uniform probability; the bound becomes $2^{-63.5}$.

Hence $(Q, J, L)$ is within $2^{-55.9}$ of $(Q, U, U')$ for independent uniform keys, and two PRF steps, on the disjoint domains $\{h_0\}$ and its complement, turn the oracle into a random function.

(2) The raw scalar of the first signature is $s_0 = a^{-1}(h_0 + r_0x) = (r_0/a)(x + c)$ with $r_0 = r(a)$, so
\[
\frac{s_0}{r_0}\,(Q - cG) = \frac{(x+c)(x-c)}{a}\,G = \frac{x^2 - c^2}{a}\,G = T .
\]
Everything on the left is public once the signature is seen, since $c = h_0/r_0$. Low-$s$ normalization may replace $s_0$ by $-s_0$ and $T$ by $-T$, which has the same abscissa. One signature therefore yields $L$, the key of every non-anchor nonce. The adversary computes $k_1 = P_L(h_1)$, reproducing retries because the validity of a candidate is publicly testable from $(Q, h_1, k_1)$, and obtains $x = (\pm k_1 s_1 - h_1)/r_1$, testing both signs against $Q$. The attack fails only if the first anchor candidate is rejected, that is if $r(a) = 0$ or $Q = -cG$; both events are publicly testable from $a = F^\sharp_x(h_0)$ and $Q$, so their probability is bounded by part~(1).
\end{proof}

\begin{remark}[scope of Theorem~\ref{thm:dsdh}]
The decisional square Diffie--Hellman assumption~\cite{bdz2003} is at least as strong as DDH, since a DDH oracle decides it, and it holds in the generic group model. It enters because the reduction must evaluate the oracle from $(Q, x^2G)$; Theorem~\ref{thm:variants} replaces it by ordinary DDH at a price listed in Table~\ref{tab:sep}. Compare $F'$ of Section~\ref{sec:counterexample}: there the PRF key is a function of $Q$ alone, so $\AIPRF$ fails; here the key $L$ depends in addition on the point $T$, which is hidden behind $x^2G$ and which one signature reveals. With $\kappa = 64$ the instance has a $2^{64}$ exhaustive-search ceiling on its $\AIPRF$ security (search for $J$ from $a$); the separation is asymptotic, with $\kappa = \lfloor\lambda/4\rfloor$, while the key recovery is concrete. We ran it on 16 secp256k1 keys with the messages \code{0}, \code{1} and a forgery on \code{2}: an independent ECDSA verifier accepted all 16 forgeries, covering the four combinations of low-$s$ sign changes, and the oracle computed from $(Q, x^2G)$ alone agreed with the signer on every tested key. [T][S]
\end{remark}

\begin{theorem}[ordering bit: hashed messages]\label{thm:order}
Let $P$ be a PRF with outputs in $\Z_n^*$ and a key $z$ independent of $x$, and fix three public messages $m_0, m_1, m_2$ with distinct digests $h_0, h_1, h_2$; for SHA-256 the one-byte messages \code{0}, \code{1}, \code{2} qualify. Let $a = P_z(h_0)$, $u = P_z(h_1)$, $v = P_z(h_2)$, $s_0 = a^{-1}(h_0 + r(a)x)$ and
\[
\beta(s) = \mathbf{1}\big\{\min([s]_n, [-s]_n) \le \lfloor n/4 \rfloor\big\}.
\]
Define $F^{\ddagger}_{(x,z)}(h_0) = a$; let $\big(F^{\ddagger}(h_1), F^{\ddagger}(h_2)\big)$ be $(u,v)$ or $(v,u)$, whichever makes $r(F^{\ddagger}(h_1)) < r(F^{\ddagger}(h_2))$ equivalent to $\beta(s_0) = 1$, with ties in $r$ broken by the scalar and the pair left as $(u,v)$ if $r(a) = 0$ or $s_0 = 0$; and $F^{\ddagger}(h) = P_z(h)$ for every other $h$. On secp256k1:
\begin{enumerate}[leftmargin=*,itemsep=1pt]
\item for every distinguisher $D$, making any number of adaptive queries,
\[
\Adv^{\AIPRF}_{F^\ddagger}(D) \le \Adv^{\PRF}_{P}(B) + 2\,\Adv^{\beta}(C) + O(1/n),
\]
where $\Adv^{\beta}(C)$ is the advantage over $1/2$ of $C$ in predicting $\beta(y)$ from $yG$ for uniform $y$; a predictor with advantage $\varepsilon$ yields a DLP solver running in time $\mathrm{poly}(\log n, 1/\varepsilon)$;
\item the distinguisher $D^\ddagger$ that requests signatures on $m_0, m_1, m_2$ and outputs~1 iff $\mathbf{1}\{r_1 < r_2\} = \beta(s_0)$ has $\Adv_{\Astar}(D^\ddagger) \ge 1/2 - O(1/n)$, in the hashed-message game and against a signer that normalizes to low $s$.
\end{enumerate}
[T]
\end{theorem}
\begin{proof}
(1) Replace $P_z$ by a random function; the PRF distinguisher $B$ samples $x$ itself. The view of $D$ is now determined by $Q$, $a$, the unordered pair $\{u,v\}$, the bit $\beta(s_0)$ and independent uniform values. If $\beta(s_0)$ is replaced by an independent fair bit, the pair $(F^\ddagger(h_1), F^\ddagger(h_2))$ is exactly uniform, because arranging two independent uniform draws in an order chosen by a fair coin restores their joint distribution. One bit is involved, so a distinguisher yields a predictor of $\beta(s_0)$ from $(Q, a)$ with half the advantage, and no hybrid over a shared secret is needed (compare Theorem~\ref{thm:hnp}).

Given a challenge $Y = yG$ with uniform $y$, sample $a$ uniformly with $r(a) \ne 0$ and set $Q = r(a)^{-1}(aY - h_0G)$. For fixed $a$ the map $y \mapsto x = (ay - h_0)/r(a)$ is an affine bijection of $\Z_n$, so $(Q, a)$ is distributed as in the game up to the event $x = 0$, and $s_0 = y$ \emph{exactly}; fixing the digest $h_0$ in advance is what makes this embedding possible. A predictor of $\beta(s_0)$ from $(Q,a)$ therefore predicts $\beta(y)$ from $Y$.

Finally, $n = 4\nu + 1$ on secp256k1, and $\beta(y) = \mathbf{1}\{[y + \nu]_n \le 2\nu\} = 1 - \lsb([2(y+\nu)]_n)$. For a uniform point $W = wG$ put $Y = 2^{-1}W - \nu G$; then $\beta(y) = 1 - \lsb(w)$. Applying the predictor to $cW$ for chosen $c$ gives an oracle for $\lsb([cw]_n)$ that is correct with probability $1/2 + \varepsilon$ over uniform $c$, and Lemma~\ref{lem:acgs} with $c_0 = 0$ recovers $w$.

(2) $\beta(-s) = \beta(s)$ and normalization does not change $r$, so the test is unaffected by it. In the real game the relation holds by construction unless a candidate is rejected, probability $O(1/n)$. In the ideal game the nonces at $h_1$ and $h_2$ are independent, uniform, and independent of the first signature, so $\mathbf{1}\{r_1 < r_2\}$ is a fair coin up to ties, which have probability at most $4/(n-1)$.
\end{proof}

\begin{remark}[$\Astar$ is not necessary]\label{rem:notnec}
Det-ECDSA with $F^\ddagger$ is as unforgeable as Rand-ECDSA, although it violates $\Astar$. After the PRF step, a forger against Rand-ECDSA simulates the signer by resampling, as in Theorem~\ref{thm:grind}: whenever a signature on one of $m_0, m_1, m_2$ is requested, it draws fresh signatures on that message until the relation $\mathbf{1}\{r_1 < r_2\} = \beta(s_0)$ is consistent with those already issued (when $\beta(s_0)$ is known and $m_1$ is asked before $m_2$, it keeps the smaller or larger $r$ of two draws). The expected number of extra queries is constant and all of them are on messages the adversary asked for, so its forgery stays fresh. Theorems~\ref{thm:planted} and~\ref{thm:dsdh}, in contrast, destroy the scheme. [T]
\end{remark}

\paragraph{Trading the assumption against setup, secrets and queries.}
Theorem~\ref{thm:dsdh} uses the square $x^2G$ because the hidden point must be computable from $x$ and from nothing else. Each of the three variants below gives up one feature of that theorem and weakens the assumption in return. Write $E(V) = \lfloor 2^\kappa x(V)/p \rfloor$ with $E(\mathcal{O}) = 0$, and $r' = r(a)$ if $r(a) \ne 0$, $r' = 1$ otherwise.

\begin{theorem}[variants of the hidden seed]\label{thm:variants}
With the notation of Theorem~\ref{thm:dsdh}, on secp256k1, each nonce function below satisfies $\AIPRF$ against any number of adaptive queries on arbitrary digests, and its key is recovered from signatures on ordinary hashed messages, also against a signer that normalizes to low $s$.
\begin{enumerate}[leftmargin=*,itemsep=3pt]
\item[(a)] \emph{A public point; DDH; two signatures.} Let $B$ be a uniform non-identity point, sampled independently of $x$ and published. Put $W = xB$, $J = E(W)$, $a = P_J(h_0)$, $T = a^{-1}(h_0B + r'W)$ and $L = \mathrm{msb}_\kappa(x(Q) \odot x(T))$; the nonce is $P_J(h)$ at $h_0$ and $P_L(h)$ elsewhere. With the probability taken also over $B$,
\[
\Adv^{\AIPRF}(D) \le \Adv^{\mathrm{DDH}}(B_0) + \Adv^{\PRF}_{P}(B_1) + \Adv^{\PRF}_{P}(B_2) + 2^{-55.9},
\]
and $T = s_0B$, so the signatures on $m_0$ and on one further message reveal $x$ as in Theorem~\ref{thm:dsdh}.
\item[(b)] \emph{No setup, one key; DDH and an $\AIPRF$; about $4\lambda$ signatures.} Let $A$ be a nonce function with outputs in $\Z_n^*$ satisfying $\AIPRF$ at $Q$. From outputs of $A_x$ at reserved inputs derive scalars $a$ and $d$, $\lambda^2$ coin bits (the parities of fresh outputs) and $2\lambda^2$ further scalars. Put $B = dG$ and $L = E\big(d\,a^{-1}(h_0G + r'Q)\big)$. Fix $2\lambda^2$ public carrier messages with distinct digests, different from $h_0$, arranged in $\lambda$ blocks of $\lambda$ pairs. Arrange the coins in $\lambda$ words of $\lambda$ bits and replace the first word that is the abscissa of a curve point by $x(B)$. The nonce is $a$ at $h_0$; on the $j$-th pair of block $i$ it is the two scalars reserved for that pair, in increasing order of $(r(k), k)$ if bit $j$ of word $i$ is $0$ and in decreasing order otherwise; and $P_L(h)$ at every other digest, the reserved inputs included. Then
\[
\Adv^{\AIPRF}(D) \le \Adv^{\AIPRF}_{A}(B_0) + \Adv^{\mathrm{DDH}}(B_1) + \Adv^{\PRF}_{P}(B_2) + 2^{-55.9} + O(1/n).
\]
An adversary reads $x(B)$ from the order of the values $r$ on the carrier messages, block by block up to the first valid word, and continues as in~(a) with $L = E(s_0B)$. It asks for $2 + 2\lambda/\rho \approx 1026$ signatures in expectation, where $\rho = (n-1)/2^{\lambda+1}$, and for at most $2 + 2\lambda^2$.
\item[(c)] \emph{Two secrets; an $\AIPRF$ alone; two signatures.} Let the signer hold, besides $x$, an independent uniform $a \in \Z_n^*$, and let $A_y$ satisfy $\AIPRF$ at $yG$. Put $s = a^{-1}(h_0 + r(a)x)$; the nonce is $a$ at $h_0$ and $A_s(h)$ elsewhere (with a fixed fallback if $r(a) = 0$ or $s = 0$). Then $\Adv^{\AIPRF}(D) \le \Adv^{\AIPRF}_{A}(B_0) + O(1/n)$, and the first signature reveals $\pm s$, the key of every later nonce.
\end{enumerate}
[T]
\end{theorem}
\begin{proof}
(a) The oracle is a function of the three points $(Q, B, W)$, and ordinary DDH~\cite{boneh1998} replaces $W = xB$ by a uniform point independent of $Q$ and $B$; this is where the independent point removes the square. Given $B$ and $J$ the map $W \mapsto T$ is an affine bijection, because $r' \ne 0$, and the statistical argument of Theorem~\ref{thm:dsdh} applies unchanged. For a valid anchor, $h_0B + r(a)W = (h_0 + r(a)x)B$, so $T = s_0B$, and its abscissa is unaffected by the sign of $s_0$. The bound is an average over $B$: for a point whose discrete logarithm $\beta$ the adversary knows, $W = \beta Q$ and the function is not an $\AIPRF$.

(b) First replace $A_x$ by a random function. The reduction to the $\AIPRF$ security of $A$ learns $a$, $d$, the coins and the reserved scalars from its oracle and computes $d\,a^{-1}(h_0G + r'Q)$ from $Q$, so it answers every query without $x$. Next, $d\,a^{-1}(h_0G + r'Q) = a^{-1}(h_0B + r'\,xB)$ with $B = dG$ uniform, and apart from this point $d$ is used only through $x(B)$; a DDH challenge $(Q, B, W)$ therefore replaces $xB$ by a uniform $W$. The hidden point is then uniform and independent of $(Q, B, a)$, its bucket $L$ is within $2^{-55.9}$ of uniform, and a PRF step makes all outputs outside $h_0$ and the carriers random. Finally, every abscissa is that of exactly two points, so $x(B)$ is uniform on the set of valid words, which is the conditional distribution of the word it replaces given its position and all other words: the replacement preserves the joint distribution of the words, the no-valid-word case included. The bits are exactly fair, since $n - 1$ is even, and independent of the reserved scalars, and ordering two independent uniform scalars by a fair bit restores their joint distribution, as in the proof of Theorem~\ref{thm:order}. For the attack, the bit of a pair is $\mathbf{1}\{r > \tilde r\}$ for its two signatures; either point with the recovered abscissa serves as $B$, because $E(s_0B)$ depends on the abscissa only. A valid word fails to appear with probability $(1-\rho)^\lambda = 2^{-256}$, and ties or rejected candidates on the carriers have probability $O(\lambda^2/n)$.

(c) Given an $\AIPRF$ challenge $(Y = yG, O)$ for $A$, sample $a$ and present $Q = r(a)^{-1}(aY - h_0G)$, as in the proof of Theorem~\ref{thm:order}: $x$ is uniform up to $O(1/n)$ and $s = y$ exactly. Answer $h_0$ with $a$ and every other digest with $O$. This is the construction if $O = A_y$ and a random function if $O$ is one. The embedding uses the independence of $a$ from $x$; with $a = A_x(h_0)$ the key $s$ would be related to $x$, and that step is not proved.
\end{proof}

\begin{table}[!htb]
\centering
\caption{Key-recovery separations $\AIPRF \not\Rightarrow \Astar$. In every row the nonce adversary queries arbitrary digests; the column ``queries'' restricts the signing attacker.}\label{tab:sep}
\small
\begin{tabularx}{\textwidth}{@{}>{\raggedright\arraybackslash}p{3.0cm}>{\raggedright\arraybackslash}p{3.3cm}X>{\raggedright\arraybackslash}p{3.3cm}@{}}
\toprule
Construction & Assumption & Setup and secrets & Signing queries \\
\midrule
$F^\dagger$ (Thm.~\ref{thm:planted}) & an $\AIPRF$ exists & none; $x$ & 2, chosen digests \\
$F^\sharp$ (Thm.~\ref{thm:dsdh}) & DSDH, PRF & none; $x$ & 2, hashed messages \\
Thm.~\ref{thm:variants}(a) & DDH, PRF & one public random point; $x$ & 2, hashed messages \\
Thm.~\ref{thm:variants}(b) & DDH, an $\AIPRF$ exists & none; $x$ & $\approx 1026$, hashed messages \\
Thm.~\ref{thm:variants}(c) & an $\AIPRF$ exists & none; $x$ and a second scalar & 2, hashed messages \\
\bottomrule
\end{tabularx}
\end{table}

\begin{remark}[what remains]
A pseudorandom function exists if DDH is hard, so (a) rests on DDH alone. We checked the variants on secp256k1 with HMAC-SHA-256 as a functional stand-in for $P$ and $A$: 16 two-signature key recoveries for (a) and 16 for (c), covering the four combinations of low-$s$ sign changes, and 10 for (b), which needed 514, 1026 or 2050 signatures; every forgery on a fresh message was accepted by an independent verifier. Such runs test the constructions and say nothing about HMAC. No row of Table~\ref{tab:sep} achieves everything at once: two hashed-message signatures, the original parameters, $x$ as the only secret, and no assumption beyond DLP or the existence of an $\AIPRF$. [T][S]
\end{remark}

\paragraph{Summary.} $\Astar$ and $\AIPRF$ are incomparable. $\AIPRF$ is not sufficient: under the $\AIPRF$ function $F^\dagger$ two chosen-digest signatures reveal the key, and under $F^\sharp$, given DSDH, so do signatures on two ordinary hashed messages; Table~\ref{tab:sep} lists the variants under DDH. It is not necessary either, if one-bit HNP is hard (Section~\ref{sec:astar-not-aiprf}). $\Astar$ is sufficient (Theorem~\ref{thm:astar-suffices}) and not necessary (Remark~\ref{rem:notnec}). All of these separations concern constructed nonce functions; whether HMAC\_DRBG satisfies $\Astar$ is untouched by them.

\section{What is provable}\label{sec:provable}

\begin{theorem}[ideal compression function]\label{thm:rom}
Model the SHA-256 compression function as a random oracle $f$. For every distinguisher $D$ for $\Astar$ that makes $q_s$ signing queries and $q_f$ queries to $f$ there is an algorithm $B$, running in time $\mathrm{Time}(D) + O((q_f+q_s)\log n)$ group operations, with
\[
\Adv_{\Astar}(D) \le \Adv^{\DLPS}(B) + \frac{16\,q_s\,(q_f + 16 q_s) + q_f}{2^{256}},
\]
where $\Adv^{\DLPS}(B)$ is the probability that $B$ computes $x$ from $Q$ given a Rand-ECDSA signing oracle $\Sign_x(\cdot;\mathcal{R})$. Moreover $\Adv^{\DLPS}(B) \le \Adv^{\EUF}_{\mathrm{Rand}}(B')$, and if $h = H(m)$ for a programmable random oracle $H$, $\Adv^{\DLPS}$ can be replaced by $\Adv^{\DLP}$. [T]
\end{theorem}
\begin{proof}
Without loss of generality $D$ never repeats $h$, since both oracles in Assumption~\ref{ass:astar} are functions of $h$; rejections in step H.3 are handled identically in all games. Each signing query costs at most 15 compression evaluations plus block~1 of step~d, which is shared. Let $\mathsf{Ask}$ be the event that $D$ queries $f$ at an input whose last 31 bytes equal $x_{[0..30]}$ (Section~\ref{sec:whitebox}). Game~0 is the real game; Game~1 aborts on $\mathsf{Ask}$. Conditioned on $\neg\mathsf{Ask}$, block~1 of steps d and f is evaluated at points $D$ never queries, so the chaining values they produce are uniform and hidden, and every later evaluation is at a fresh point unless an internal collision occurs or $D$ queries a hidden chaining value; this has probability at most $(16 q_s(q_f + 16q_s) + q_f)/2^{256}$. In Game~2 the nonces are therefore independent and uniform, which is the $\Sign_x(\cdot;\mathcal{R})$ oracle. The games are identical until $\mathsf{Ask}$ or a collision, so it remains to bound $\Pr[\mathsf{Ask}]$ in Game~2. There $B$ receives $Q$ and the signing oracle, answers signing queries with it and $f$-queries by lazy sampling, and keeps the table $T = \{Q - bG : 0 \le b \le 255\}$. For every $f$-query with tail $w$ it checks whether $2^8wG \in T$, at the cost of one scalar multiplication, and on a hit outputs $x = 2^8w + b$. No query index is guessed, so there is no factor $q_f$. $B'$ runs $B$ and forges with the recovered $x$. If $H$ is programmable, the simulation~\eqref{eq:sim} answers signing queries without $x$.
\end{proof}

\begin{remark}
The distinguisher in Assumption~\ref{ass:astar} chooses $h$ itself. Without a programmable message hash, a DLP reduction cannot answer its signing queries (Proposition~\ref{prop:aiprf-insufficient}), so the ideal-compression-function argument yields $\Astar$ from DLP in the presence of a signing oracle, not from plain DLP. Combined with Theorem~\ref{thm:astar-suffices}, $\Adv^{\EUF}_{\mathrm{Det}} \le 2\,\Adv^{\EUF}_{\mathrm{Rand}} + (16q_s(q_f+16q_s)+q_f)/2^{256}$ up to the running-time overhead of $B$.
\end{remark}

\begin{theorem}[independent nonce key]\label{thm:detprime}
For Det$'$-ECDSA with $k = P_z(h)$ and $z$ independent of $x$ (or $(x,z) = \mathrm{PRG}(\mathrm{seed})$),
\[
\Adv^{\EUF}_{\mathrm{Det}'}(A) \le \Adv^{\EUF}_{\mathrm{Rand}}(B) + \Adv^{\PRF}_{P}(D)
\]
in the standard model. [T]
\end{theorem}
\begin{proof}
The $\PRF$ distinguisher $D$ knows $x$ (it is not the PRF key) and simulates signatures with the nonces returned by its oracle.
\end{proof}

This is precisely the construction that RFC~6979~\S3.6 describes as ``additional secret data \dots may help in formally proving the security''. EdDSA~\cite{rfc8032} obtains it by splitting $(a, \mathrm{prefix}) = \mathrm{SHA\text{-}512}(\mathrm{seed})$. BIP-340~\cite{bip340} does not: $t = d \oplus H_{\mathrm{aux}}(\mathrm{aux})$, $k = H_{\mathrm{nonce}}(t \,\|\, P \,\|\, m)$ still needs an $\Astar$-type assumption, discharged in the ROM. Table~\ref{tab:summary} summarizes.

\begin{table}[t]
\centering
\caption{Assumptions beyond the security of Rand-ECDSA.}\label{tab:summary}
\small
\begin{tabularx}{\textwidth}{@{}>{\raggedright\arraybackslash}p{4.1cm}X>{\raggedright\arraybackslash}p{2.6cm}>{\raggedright\arraybackslash}p{2.9cm}@{}}
\toprule
Scheme & Assumption & Model & Status \\
\midrule
$k = \mathrm{HMAC\_DRBG}(x\|h)$, RFC 6979 & $\Astar$ & standard & open; no attack; falsifiable \\
same & DLP with a Rand-ECDSA signing oracle; plain DLP if $H$ programmable & ideal compression function & proved, no $q_H$ factor \\
$k = \mathrm{HMAC\_DRBG}(x\|h\|k')$, RFC~6979 \S3.6 & $\Astar$ for any $k'$; nothing extra if $k'$ uniform and secret & ROM & as above \\
BIP-340 Schnorr & $\Astar$-analogue for tagged SHA-256 & ROM & proved in ROM \\
$k = P_z(h)$, $z$ independent & PRF security of $P$ & standard & proved \\
EdDSA prefix & PRF security of $H$ with prefix as key & standard (ROM in~\cite{bcjz2021}) & proved in ROM \\
$F'(x,h) = P_{\Ext(xG)}(h)$ & --- & standard & PRF, but key revealed by one signature \\
RFC~6979 over $f^*$ (Section~\ref{sec:whitebox}) & --- & standard; $f^*$ has (A1), (A2), (A2$'_{248}$) given CR & PRF in $x$, key revealed by one signature \\
$F^\dagger$, planted collision (Theorem~\ref{thm:planted}) & --- & standard & $\AIPRF$, key revealed by two chosen-digest signatures \\
$F^\sharp$, hidden seed (Theorem~\ref{thm:dsdh}) & --- & standard & $\AIPRF$ under DSDH + PRF; key revealed by signatures on two hashed messages \\
Variants of $F^\sharp$ (Theorem~\ref{thm:variants}) & --- & standard & $\AIPRF$ under DDH with a public point, or DDH and an $\AIPRF$, or an $\AIPRF$ and a second secret; key revealed through hashed messages \\
$F^\ddagger$, ordering bit (Theorem~\ref{thm:order}) & --- & standard & $\AIPRF$ under PRF + DLP; $\Astar$ fails with advantage $1/2$; unforgeable \\
\bottomrule
\end{tabularx}
\end{table}

\section{Curve dependence of the standard part}\label{sec:curves}

$\Astar$ itself is curve-independent. What depends on the curve is the standard part of a proof that HMAC\_DRBG is a PRF in $x$. In step~d of RFC~6979~\S3.2 the HMAC key is the public $K_0 = \mathtt{0x00}^{32}$, and $x$ enters the \emph{message}: bytes 0--30 fill the tail of block~1 and byte~31 opens block~2 (Section~\ref{sec:whitebox}). Bellare's PRF theorem for HMAC~\cite{bellare2006}, which assumes a uniform secret key, therefore does not apply to step~d. What step~d needs is property (A2$'_{248}$): $f(IV_1, V_0\|\mathtt{0x00}\|x_{[0..30]})$ is pseudorandom. Once $K_1$ is pseudorandom, Bellare's theorem applies to $\mathrm{HMAC}_{K_1}$ in the later steps. For $x$ uniform on $[1,n-1]$, the secret tail $x_{[0..30]} = \lfloor x/256 \rfloor$ is at statistical distance $\approx (2^{256}-n)/2^{256}$ from uniform on $\bits^{248}$.

\begin{table}[h]
\centering
\caption{Distance of the secret tail $x_{[0..30]}$ of block~1 from uniform. [T]}\label{tab:curves}
\begin{tabular}{@{}lll@{}}
\toprule
Curve & Distance & Consequence \\
\midrule
secp256k1 & $2^{-127.65}$ & (A2$'_{248}$) applies with loss $2^{-127.65}$ \\
P-256 & $2^{-32}$ & (A2$'_{248}$) applies with loss $2^{-32}$ \\
brainpoolP256r1 & $\approx 0.336$ & extra assumption: (A2$'_{248}$) for keys from a density-$0.664$ subset \\
\bottomrule
\end{tabular}
\end{table}

The same number governs every place in the secp256k1 stack where a 256-bit value is reduced modulo $n$ without rejection. For $T$ uniform on $[0,2^{256})$ and $k = T \bmod n$, the statistical distance from uniform, the Fourier coefficient $|B_1| = |\mathbb{E}\, e^{2\pi i k/n}|$ relevant to Bleichenbacher's attack, and $\Pr[T \ge n]$ all equal $1.2708 \cdot 2^{-128} = 2^{-127.654}$ to three digits~[T]. A data threshold of order $|B_1|^{-2} \approx 2^{255}$ signatures makes the bias irrelevant in practice; transplanted verbatim to P-256 the same operation gives $2^{-32}$ (threshold $\approx 2^{64}$) and on brainpoolP256r1 $0.166$ (lattice attacks with hundreds of signatures). RFC~9591 (FROST) requires $|n - 2^{256}| < 2^{128}$ to skip the range check; secp256k1 misses this criterion by $0.35$ bit, which changes nothing in practice. [T]

\section{Deployed implementations}\label{sec:impl}

We reviewed the source of 31 secp256k1 implementations (18 libraries, 7 wallets and nodes, 6 hardware wallets) at their default branches on 2026-10-18, and libsecp256k1 at commit \code{04b91ce} (2026-10-19). Line numbers refer to those snapshots. [S]

\subsection{Nonce defaults}

28 of 31 implementations (90\%) sign ECDSA deterministically without additional entropy; none of the 18 libraries hedges by default although 9 expose an API for it. The exceptions are python-ecdsa (\code{sign()} draws a uniform $k$), LDK (hedges its own commitment and sweep transactions) and BitBox02 (anti-klepto with host entropy). For BIP-340, 15 of 26 implementations sign with zero auxiliary randomness; Bitcoin Core's source reads ``Use \code{uint256\{\}} as \code{aux\_rnd} for now''. The RFC~6979~\S3.6 construction with an independent nonce key, which would discharge $\Astar$ by Theorem~\ref{thm:detprime}, is used nowhere in the sample. [S]

Among general-purpose libraries the default differs: OpenSSL~3.x hedges (\code{ossl\_bn\_gen\_\allowbreak dsa\_nonce\_\allowbreak fixed\_top}, SHA-512 over $\mathrm{priv}\,\|\,\mathrm{dgst}\,\|\,\mathrm{random}$; RFC~6979 only with \code{nonce\_type=1}); wolfSSL, Crypto++, libgcrypt, Botan (current head), Bouncy Castle's stock \code{ECDSASigner} and python-ecdsa draw random nonces, with RFC~6979 as an opt-in; fastecdsa is deterministic by default; MIRACL Core has no RFC~6979 at all. None of these produces a biased nonce out of the box: the random paths use either rejection sampling or at least 64 extra bits before reduction. [S]

\subsection{Compliance: \texorpdfstring{\code{bits2octets}}{bits2octets}}

RFC~6979 feeds $\mathrm{bits2octets}(h) = \mathrm{int2octets}(\mathrm{bits2int}(h) \bmod n)$ to the DRBG. libsecp256k1 reduced the message only from version 0.2.0 (2022-12-12); the changelog reads ``now reduces the message hash modulo the group order to match the specification''. Ports made earlier retain the raw hash: NBitcoin (\code{msg32.CopyTo(keydata)}), dcrd (\code{copy(keyBuf[offset:],} \code{hash)}) and btcec~v2, which delegates to dcrd; through them LND, btcd and go-ethereum in its non-cgo build. go-ethereum's cgo build uses a vendored libsecp256k1 that reduces, so the two builds of the same program produce different signatures when $h \ge n$. On secp256k1 with SHA-256 the discrepancy is observable only when $h \ge n$, probability $2^{-127.65}$, and it has no security consequence: $h$ is public and $\Astar$ is stated for any injective encoding of $h$. It is a compatibility bug. [S][T]

elliptic and Bouncy Castle truncate to $\lceil \log_2 n \rceil$ bits before a single subtraction of $n$; since the truncated value is below $2^{\lceil \log_2 n\rceil} \le 2n$, one subtraction is exactly \code{bits2octets} for every hash length and every curve. The Ledger SDK's open implementation (\code{lib\_cxng/}\allowbreak\code{src/}\allowbreak\code{cx\_rng\_rfc6979.c}) has two $2^{-256}$-probability deviations: it leaves $\mathrm{bits2int}(h)$ unreduced when it equals $n$ exactly, and it does not reject the candidate $k = 0$. [S][T]

\subsection{Retry and grinding}

All implementations with an $F_x$ advance the DRBG on rejection as in step~H.3. Trezor and the Ledger SDK apply the $K,V$ update eagerly after every output, libsecp256k1 lazily at the start of the next \code{generate}; the output sequences coincide. [S]

Bitcoin Core (default since 0.17), libwally (\code{EC\_FLAG\_GRIND\_R}, used by Core Lightning), rust-secp256k1 (\code{sign\_ecdsa\_low\_r}, used by BDK), Jade and Coldcard grind for a low $R$: they pass a counter $c$ as the \S3.6 additional data (no data for the first attempt, then the 32-byte little-endian counter) and keep the first signature whose $r$ fits in 32 DER bytes. This selects $k = F_x(h,c^*)$ with $c^* = \min\{c : \pi(F_x(h,c)\,G)\}$ for the public predicate $\pi(R) \Leftrightarrow x(R) \bmod n < 2^{255}$.

\begin{theorem}[grinding]\label{thm:grind}
Let $\pi$ be any public predicate on points with $\rho = \Pr_k[\pi(kG)] > 0$. For every forger $A$ against grinding Det-ECDSA there are $B$ and $D$ with
\[
\Adv^{\EUF}_{\mathrm{Grind\text{-}Det}}(A) \le \Adv^{\EUF}_{\mathrm{Rand}}(B) + \Adv_{\Astar_c}(D),
\]
where $\Astar_c$ is $\Astar$ for $F_x(h,c)$ on the extended input domain and $B$ makes $q_s/\rho$ signing queries in expectation. [T]
\end{theorem}
\begin{proof}
By $\Astar_c$ the candidates $k_0, k_1, \dots$ are independent and uniform, so the selected nonce is uniform on $S_\pi = \{k : \pi(kG)\}$. $B$ simulates this signer with the Rand oracle: on query $m$ it requests signatures on $m$ until one satisfies $\pi$, returns it, and discards the rest. The distribution is identical (first success in a sequence of independent uniform nonces). Every extra query of $B$ is on a message $A$ queried, so $A$'s forgery is fresh for $B$. The number of attempts is geometric with parameter $\rho$; for low $R$, $\rho \approx 1/2$. The attempt count $c^*$, observable through timing, adds nothing: $B$ reproduces its distribution.
\end{proof}

$\Astar_c$ is the assumption already required by the rejection loop, which also selects a non-first output. Under it the selected nonce is uniform on the low-$R$ set, whose Fourier coefficients---the bias a Bleichenbacher-type attack would use---can be bounded without any model.

\begin{proposition}[the low-$R$ set]\label{prop:lowr}
Let $L = 2^{255}$, $e_m(w) = \exp(2\pi i w/m)$, $S = \{k \in \Z_n^* : x(kG) < L\}$ and $S' = \{k \in \Z_n^* : x(kG) \bmod n < L\}$, the set selected by grinding. On secp256k1, with its standard generator,
\[
\Big|\frac{|S|}{n-1} - \frac12\Big| < 2^{-118.9},
\qquad
\max_{t \not\equiv 0}\ \Big|\frac{1}{|S|}\sum_{k \in S} e_n(tk)\Big| < 2^{-117.9},
\]
and the same holds for $S'$ with $2^{-117.9}$ and $2^{-116.9}$. Both sets are symmetric under $k \mapsto -k$, so all coefficients are real. [T][L]
\end{proposition}
\begin{proof}
Expand the indicator of $[0,L) \subset \mathbb{F}_p$ in additive characters, $\mathbf{1}_{[0,L)}(w) = \sum_j c_j\, e_p(jw)$. Then $c_0 = L/p$, and $|c_j| \le 1/(p\,|\sin(\pi j/p)|) \le 1/(2|j|)$ for $0 < |j| \le (p-1)/2$ gives $\sum_{j \ne 0} |c_j| \le 1 + \ln((p-1)/2)$. The group $E(\mathbb{F}_p)$ is cyclic of prime order $n$, so its characters are $kG \mapsto e_n(tk)$, and the function $x$ has degree~2 with a single pole, at infinity. The bound of Kohel and Shparlinski~\cite{ks2000} for functions whose polar divisor is supported at one point gives, for every $j \not\equiv 0$ and every $t$,
\[
\Big|\sum_{k=1}^{n-1} e_n(tk)\, e_p\big(j\,x(kG)\big)\Big| \le 3\sqrt{p}.
\]
Put $D = 3\sqrt{p}\,\big(1 + \ln((p-1)/2)\big) < 2^{137.06}$. The term $j = 0$ contributes $(n-1)L/p$ for $t = 0$ and $-L/p$ for $t \not\equiv 0$, so
\[
\big|\,|S| - (n-1)L/p\,\big| \le D,
\qquad
\Big|\sum_{k \in S} e_n(tk) + L/p\Big| \le D \quad (t \not\equiv 0),
\]
and $L/p - 1/2 = 2^{-225}$. For $S'$ the set of abscissas is $[0,L) \cup [n,p)$, since $p - n < L$: two intervals, hence the error $2D$, around the density $(L + p - n)/p = 1/2 + 2^{-127.65}$.
\end{proof}

An attack driven by this bias would need more than $2^{233}$ signatures. The bound is unconditional and holds for the actual curve and generator; it is within a factor of about $2^{11}$ of the $2^{-128}$ typical of a random subset of density $1/2$, which is what the bijective random-oracle model of~\cite{fkp2016} would predict, and the difference is the logarithmic loss of the completion argument. On a toy prime-order curve with $p = 65167$ the largest coefficient we measured is $2.2\sqrt{p}$ before normalization, against the bound $34\sqrt{p}$. [T] The cost of grinding is an extra loop in the signer (a fault-injection surface) and twice the $F_x$ evaluations.

\subsection{Testability}

The branch $k \notin [1,n-1]$ is unreachable with natural inputs. libsecp256k1 tests it with \code{nonce\_function\_test\_retry} (\code{tests.c:6721--6744}), which returns $0$, $2^{256}-1$, $2^{256}-2$, $n$ and $n+1$ before a valid candidate and checks that the result equals the untampered RFC~6979 output. The branch $h \ge n$ is tested by no implementation we examined; Appendix~\ref{app:vectors} gives vectors that distinguish the conformant and raw-hash variants. Deterministic and hedged signatures are indistinguishable on chain, so the deployed share of each cannot be measured. [S]

\section{Nonce bias in context}\label{sec:bias}

A single error-free leaked most-significant bit of $k$ on a 256-bit curve has not been attacked. Aranha et al.~\cite{aranha2014} broke a one-bit bias on a 160-bit curve with $2^{33}$ signatures and estimated $2^{52}$ signatures, $2^{58}$ time, $2^{52}$ memory for 256 bits by the scaling $2^{N/5}$. LadderLeak~\cite{ladderleak2020} broke one-bit leakage on P-192 with about $2^{29}$ signatures and left P-256 with one-bit leakage open; its P-256 estimate ($\approx 5 \cdot 10^5$ signatures) is for two-bit leakage. Lattice methods reach one-bit leakage on 160-bit curves~\cite{gao2024} and two-bit leakage on 256-bit curves with 146 signatures~\cite{xu2023}.

Our model of the sort-and-difference/FFT attack (five reduction rounds, 24-bit FFT) gives $2^{47.4}$ signatures for one error-free bit on 256 bits. A toy calibration on curves $q = 2^m - 1$ for $32 \le m \le 88$ reproduces the model within $1.3$ bits (mean absolute gap $0.3$ bit), and calibration against the LadderLeak P-192 experiment gives $2^{40}$--$2^{45}$. The honest envelope is $2^{40}$--$2^{47}$; the figure $2^{52}$ is the published scaling. [O] These are to be contrasted with $2^{255}$ for the naive-reduction bias of Section~\ref{sec:curves}: the two regimes differ by 127 bits in $|B_1|$ and are not continuations of each other.

GLV decomposition in the signing path (noble-secp256k1~1.7.1, dcrd under the \code{tinygo} tag, k256 without \code{precomputed-tables}, elliptic's fallback) decomposes a uniform $k$ into $k = k_1 + \lambda k_2$ and leaves the distribution of $k$ unchanged; the dangerous pattern is \emph{recomposition} from short random halves, which no reviewed implementation does, and which on secp256k1 ($D = -3$) yields a short-vector bias of $0.112$ rather than the GLS value $2/\pi$ (exact computation; Monte Carlo $0.113$). The real GLV-specific risk is the decomposition side channel of~\cite{aranha2014}, a lattice attack with tens of signatures. [S][T]

\section{Conclusion and open problems}

The security of RFC~6979 on secp256k1 is equivalent, up to the security of randomized ECDSA, to the assumption $\Astar$ on HMAC\_DRBG at known public key. $\Astar$ does not follow from the PRF security of HMAC and the hardness of DLP, nor, given a collision-resistance premise, from the PRF and dual-PRF properties of the SHA-256 compression function. It is not implied by auxiliary-input PRF security either: in the standard model there are nonce functions with that property under which two signatures reveal the key---on chosen digests if any such function exists, on ordinary hashed messages under decisional square Diffie--Hellman, or under DDH with one public random point---or under which three signatures on fixed messages distinguish the transcript of a scheme that stays unforgeable. $\Astar$ follows, with an ideal compression function, from DLP in the presence of a randomized-ECDSA signing oracle; for HMAC\_DRBG itself it is violated by no known attack, and it is falsifiable. It is probably the most widely deployed unproven assumption in ECDSA practice. Separating the nonce key from the signing key, as RFC~6979~\S3.6 and EdDSA do, makes it unnecessary, but is incompatible with existing keys and is used by none of the 31 implementations audited.

Open: a standard-model proof or refutation of $\Astar$ for HMAC\_DRBG; removing the collision-resistance premise from Theorem~\ref{thm:whitebox}; a two-signature key-recovery separation through ordinary message hashing with the original parameters and $x$ as the only secret, from DLP or from the bare existence of an $\AIPRF$ (Table~\ref{tab:sep} lists what each relaxation costs); closing the factor $2^{11}$ between the proved and the heuristic Fourier bound for the low-$R$ set; and a measured, rather than modelled, cost for one-bit nonce leakage on a 256-bit curve.

\appendix

\section{Lemma~\ref{lem:acgs}: details}\label{app:acgs}

Write $[z]_n$ for the residue of $z$ in $\{0,\dots,n-1\}$, $n$ an odd prime. Let $t_k = 2^{-k} \bmod n$, $K = \lceil \log_2 n \rceil$, $m = \lceil 8\log_2 n/\varepsilon^2 \rceil$, $M = \lceil 8/\varepsilon \rceil$. Fix random $(u,v) \in \Z_n^2$ and query $g$ at $a_{k,i} = t_k + iu + v$ for $k = 0,\dots,K$, $i = 1,\dots,m$. Then
\[
a_{k,i}x + c_0 \equiv t_k x + iY + Z \pmod n, \qquad Y = [ux]_n,\ Z = [vx + c_0]_n .
\]
Enumerate $\beta_Y = \lsb(Y)$, $\beta_Z = \lsb(Z)$, an approximation $\alpha_Y = U/(mM)$ with $0 \le Y/n - \alpha_Y < 1/(mM)$, an approximation $\alpha_Z = V/M$ with $0 \le Z/n - \alpha_Z < 1/M$, and an initial approximation $\gamma_0 = X_0/M$ with $0 \le [t_0x]_n/n - \gamma_0 < 1/M$; in total $4mM^3 = O(\varepsilon^{-5}\log n)$ branches, and the query set does not depend on the branch.

In the correct branch maintain $0 \le [t_kx]_n/n - \gamma_k < 1/M$. For $S_{k,i} = ([t_kx]_n + iY + Z)/n$ and $\hat S_{k,i} = \gamma_k + i\alpha_Y + \alpha_Z$ one has $\hat S_{k,i} \le S_{k,i} < \hat S_{k,i} + 3/M$, hence $\lfloor S_{k,i} \rfloor = \lfloor \hat S_{k,i} \rfloor =: w_{k,i}$ whenever $[a_{k,i}x + c_0]_n \ge 3n/M$, and then, $n$ being odd,
\[
\lsb([t_kx]_n) = g(a_{k,i}) \oplus (i\beta_Y \bmod 2) \oplus \beta_Z \oplus (w_{k,i} \bmod 2)
\]
whenever $g$ answers correctly. The event ``$g(a)$ correct and $[ax + c_0]_n \ge 3n/M$'' depends only on $a$, has probability $\ge \tfrac12 + \varepsilon - 3/M \ge \tfrac12 + \varepsilon/2$, and the indicators over $i$ are pairwise independent because $a_{k,i}$ are pairwise independent and uniform ($n$ prime, $m < n$). Chebyshev bounds the majority-vote error at step $k$ by $1/(m\varepsilon^2) \le 1/(8\log_2 n)$; a union bound over $K+1$ steps gives success probability $\ge 3/4$. With the bit $b_k = \lsb([t_kx]_n)$ known, $[t_{k+1}x]_n = ([t_kx]_n + b_k n)/2$, so $\gamma_{k+1} = (\gamma_k + b_k)/2$ preserves the invariant and halves the error; after $K$ steps the error is below $1/n$ and $[t_Kx]_n = \lceil n\gamma_K \rceil$ (the ceiling, since the invariant is one-sided), whence $x = 2^K [t_Kx]_n \bmod n$, verified against $xG = Q$. \qed

\section{Test vectors for the \texorpdfstring{$h \ge n$}{h >= n} branch}\label{app:vectors}

secp256k1, SHA-256, HMAC\_DRBG as in RFC~6979~\S3.2; the implementation used to produce these values reproduces the RFC's own P-256/SHA-256 vector for the message ``sample'' ($k = \mathtt{A6E3C57D\ldots8AAD60}$). ``RFC'' seeds the DRBG with $\mathrm{int2octets}(x)\,\|\,\mathrm{bits2octets}(h)$; ``raw'' seeds it with $x \,\|\, h$ unreduced (libsecp256k1 before 0.2.0, NBitcoin, dcrd, btcec~v2).

\medskip
\noindent $x = 1$, $h = n + 1$ ($h \bmod n = 1$):
{\small\begin{align*}
k_{\mathrm{RFC}} &= \mathtt{9a409dab05968059da3efb323dc67c96f234571b965fd39810ca0643fbb795ac}\\
k_{\mathrm{raw}} &= \mathtt{daf2c4e18d51a5918522bc9ab5856ed4682208ed35315c9365b86e90dae8e3e3}
\end{align*}}
$x = 1$, $h = 2^{256}-1$ ($h \bmod n = t$):
{\small\begin{align*}
k_{\mathrm{RFC}} &= \mathtt{71139aac71b52f7d5961915af1b30f94baf35e39b0043c33d41a57d476a8905c}\\
k_{\mathrm{raw}} &= \mathtt{1a3bf84f9a0fd1fff4655ba0ec976b542621fd87b88c3ca6770d4183cbc0e741}
\end{align*}}
Expected behaviour: libsecp256k1~$\ge 0.2.0$, secp256k1-zkp, elliptic, Bouncy Castle, Trezor, the Ledger SDK, OpenSSL, libgcrypt and fastecdsa return $k_{\mathrm{RFC}}$; NBitcoin, dcrd, btcec~v2, LND and go-ethereum (non-cgo) return $k_{\mathrm{raw}}$. Signatures must verify in both cases, since the signature equation uses $h \bmod n$.

\section*{Declaration of Generative AI Use}
During the preparation of this work, the authors used Claude Fable 5.2 and ChatGPT Astra to prove mathematical correctness, improve readability, check grammar, and refine phrasing. After using this tool, the authors manually reviewed and edited all text and take full responsibility for the content of the publication.

\end{document}